\documentclass[a4paper,twocolumn,13pt]{quantumarticle}
\pdfoutput=1
\usepackage[utf8]{inputenc}
\usepackage[english]{babel}
\usepackage[T1]{fontenc}
\usepackage{hyperref}
\usepackage{tikz}
\usepackage{lipsum}
\usepackage{amssymb}
\usepackage{bm}% bold math
\usepackage{lineno,hyperref}
\usepackage{colortbl}
\usepackage{amsmath,amssymb,amsfonts}
\usepackage{algorithmic}
\usepackage{graphicx}
\usepackage{epstopdf}
\usepackage{float}

\begin{document}

\title{Orthogonality catastrophe and quantum speed limit for dynamical quantum phase transition}

\author{Zheng-Rong Zhu}
\affiliation{School of Physics, Beijing Institute of Technology, Beijing 100081, China}
\author{Bin Shao}
\email{sbin610@bit.edu.cn}
\affiliation{School of Physics, Beijing Institute of Technology, Beijing 100081, China}
\author{Jian Zou}
\affiliation{School of Physics, Beijing Institute of Technology, Beijing 100081, China}
\author{Lian-Ao Wu}
\affiliation{Department of Physics, University of the Basque Country UPV/EHU, 48080 Bilbao,
	IKERBASQUE, Basque Foundation for Science, 48011 Bilbao,
	EHU Quantum Center, University of the Basque Country UPV/EHU, Leioa, Biscay 48940, Spain}

\maketitle

\begin{abstract}
  We investigate the orthogonality catastrophe and quantum speed limit in the Creutz model for dynamical quantum phase transitions. We demonstrate that exact zeros of the Loschmidt echo can exist in finite-size systems for specific discrete values. We highlight the role of the zero-energy mode when analyzing quench dynamics near the critical point. Additionally, we examine the behavior of the time for the first exact zeros of the Loschmidt echo and the corresponding quantum speed limit time as the system size increases. While the bound is not tight, it can be attributed to the scaling properties of the band gap and energy variance with respect to system size. As such, we establish a link between the orthogonality catastrophe and quantum speed limit by referencing the full form of the Loschmidt echo. In addition, we reveal the potential that the quantum speed limit holds to detect static quantum phase transition point and a reduced amplitude of the noise induced behaviors of quantum speed limit.
\end{abstract}

\section{Introduction}
The dynamics of non-equilibrium systems have gained significant attention due to the growing interest in understanding the properties of the dynamical evolution of many-body systems~~\cite{RevModPhys.83.863}. The simplest experimental protocol to drive a system out of equilibrium is through a quantum quench~~\cite{PhysRevLett.96.136801}.  In this scenario, the properties of the dynamical evolution following a sudden quench can be characterized by a measure known as the Loschmidt echo (LE)~~\cite{GORIN200633,PhysRevLett.111.046402}, which is the modulus of the Loschmidt amplitude as a measure of the dynamical overlap between the initial quantum state that is influenced by the pre-quench and post-quench Hamiltonians, respectively.  The LE, which allows one to capture the main features of a system, is defined by $\mathcal{L}(t)=|\chi(t)|^2$, 
where $\chi(t)=\langle \Psi| e^{i H_f t} e^{-i H_i t} |\Psi\rangle$ and $\hbar =1$ for brevity. The initial state $|\Psi \rangle$ is an eigenstate of the initial Hamiltonian $H_{i}$, while $H_{f}$ represents the final perturbed Hamiltonian. Furthermore, the quantity $\chi(t)$ can be used to describe the phenomenon of dynamical orthogonality, specifically the concept known as orthogonality catastrophe (OC)~\cite{PhysRevLett.18.1049,PhysRevLett.124.110601}. The concept of OC has been studied in various contexts, including quantum spin models~\cite{PhysRevB.94.014310,PhysRevE.74.031123} and trapped gases~\cite{PhysRevLett.111.165303,PhysRevLett.112.246401}, and has been found to have a connection with  the
adiabaticity breakdown~\cite{PhysRevLett.119.200401}, mimicking unknown quantum states~\cite{pyshkin2022mimicking} and the quantum speed limit (QSL)~\cite{PhysRevLett.124.110601,PhysRevA.107.042427,Deffner_2017}.

Interestingly, if we focus on the scenario where $|\Psi\rangle$ represents the ground state of $H_{i}$, then the expression for LE can be simplified as follows:
\begin{equation}
	\mathcal{L}(t)=|\langle \Psi|e^{-iH_f t}| \Psi \rangle|^2.
\end{equation}
We will now direct our attention towards the dynamical quantum phase transition (DQPT)~\cite{PhysRevLett.110.135704,Heyl_2018,PhysRevB.89.125120,zvyagin2016dynamical}, which is an indication of nonequilibrium critical phenomena. The DQPT is characterized by the occurrence of zeros of the LE at specific critical times $t_{n}^{*}$, and it has been studied in various systems, including topological~\cite{PhysRevB.91.155127,PhysRevB.96.014302} and Floquet systems~\cite{PhysRevB.105.094311,PhysRevB.105.094304}. Furthermore, these zeros of LE correspond to non-analytic behaviors of the dynamical free energy, specifically the rate function $\lambda(t)=-\frac{1}{L}\ln \mathcal{L}(t)$, at those critical times~\cite{PhysRevLett.110.135704}. In general, exact zeros of the LE or non-analyticities of the dynamical free energy only occur as the system size $L$ approaches infinity~\cite{PhysRevLett.110.135704}. However, recent studies have shed light on the fact that the exact zeros of LE can be observed for finite-sized systems~\cite{PhysRevB.104.094311,PhysRevB.107.134302,10.21468/SciPostPhys.10.5.100}, depending on the selection of post-quench parameters. This has enhanced our understanding of DQPT. Moreover, the critical time at which the first exact zeros of LE occur is identified as the minimum time required for an initial state to become orthogonal to the evolving state, which is directly related to the QSL time. This establishes a direct correlation between QSL and DQPT~\cite{PhysRevB.104.094311}. 

It is noteworthy that the exact zeros of LE provide insight into the phenomena of the OC, which captures the behavior of dynamical orthogonality as the system size increases. Moreover, there is evidence suggesting that the OC and QSL are interconnected~\cite{PhysRevLett.124.110601}. This underscores the importance of examining these phenomena in the context of DQPT.

The goal of this paper is to investigate the relationship between OC and QSL within the framework of DQPT, using the Creutz model~\cite{PhysRevB.99.054302,PhysRevLett.83.2636} as a case study. The Creutz model describes the behavior of spinless fermions hopping on a two-leg ladder in the presence of a magnetic field, and it is one of exactly solvable models ~\cite{PhysRevB.58.R1703,WU1992280}. The magnetic field induces a quantum phase transition (QPT)~\cite{PhysRevLett.96.140604,PhysRevA.82.062119} between two insulating phases with different current configurations~\cite{PhysRevLett.102.135702}. The insulating gap in the system with a commensurate number of sites can close at the quantum critical point, depending on the choice of hopping amplitudes. The finite-size quantization condition determines the wave number that defines this gap closing point. By altering the hopping amplitudes, the gap closing point and associated zero-energy modes can be fine-tuned and moved~\cite{PhysRevLett.102.135702}. It has been shown that the zero-energy modes control the revival period of the LE, which may exhibit jumps based on the system parameters~\cite{PhysRevB.99.054302,PhysRevA.85.032114,PhysRevLett.112.220401}. The role of the zero-energy modes in the DQPT is emphasized when the system is quenched to the quantum critical point. Our analysis of OC and QSL in this system will rely on the important role played by the zero-energy modes.

This paper demonstrate the existence of exact zeros of LE for specific discrete values in finite-size system, Creutz model. The role of the appearing zero-energy mode when the system size is commensurate with the quantization condition, upon quenching to the quantum critical point is emphasized. We also analyze the scenario where the zero-energy mode is absent. Furthermore, we investigate the connection between QSL and OC in the context of DQPT by studying the maximum and minimum of the first divergence time of the rate function, denoted as $\tau_{fmax}$ and $\tau_{fmin}$, respectively. We calculate the QSL time for both cases and show that $\tau_{fmax}$ tends to infinity as the system size approaches infinity, which is related to the quench dynamics near the quantum critical point where the band gap is vanishing. On the other hand, $\tau_{fmin}$ converges to a constant value due to the finite band gap. However, the QSL time approaches zero for both cases, and the bound is not tight because we scrutinize all $k$ modes and the energy variance scaling as the size of the system. Therefore, we uncover the relationship between OC and QSL by examining the complete structure of LE that encompasses all $k$ modes. We further demonstrate non-analytical behavior evident in the average and variance of $\tau_{QSL}$ when quenched across the critical point. We also reveal a comparable behaviors to the noiseless situation, with the exception of a reduced amplitude of the QSL induced by noise.

The paper is organized as follows. In Section \ref{section2}, we provide a brief review of the Creutz model and analyze the appearance of zero-energy modes. In Section \ref{section3}, we outline the exact conditions for the occurrence of zeros of LE and analyze the dynamical behaviors of the allowed quench parameter regions, especially those close to the critical point where zero-energy modes may arise. In Section \ref{section4}, we demonstrate the behaviors of the time for the first exact zeros of Loschmidt echo and the corresponding QSL time. We establish a connection between OC and QSL. In Section \ref{section5}, We reveal the behavior of QSL in presence of noise. Finally, we conclude in Section \ref{section6}.

\begin{figure}[ht]
	\vspace{-1.6cm}
	\includegraphics[width=1\linewidth]{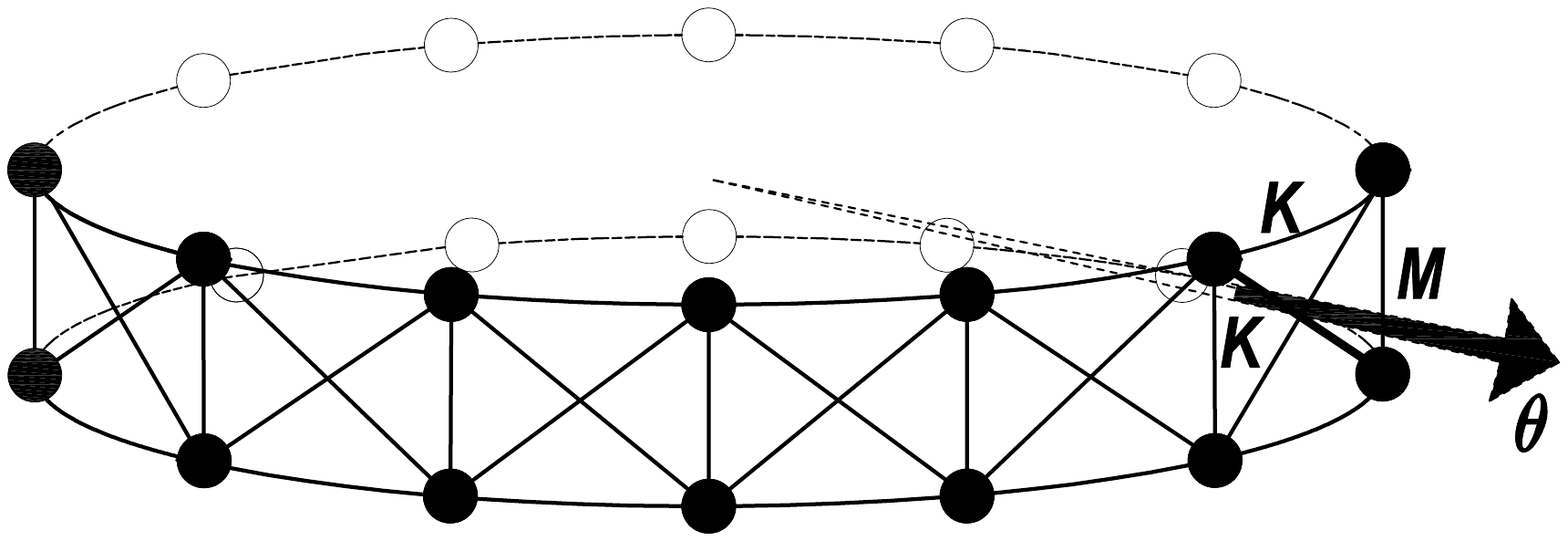}
	\vspace{-1.6cm}
	\caption{The diagram of Creutz ladder with periodic boundary conditions and magnetic flux $\theta$. The hopping amplitudes of horizontal and diagonal bonds are $K$, and vertical is $M$.  }
	\label{Figure1}
\end{figure}	
%\bigskip
\section{Creutz Model\label{section2}} 
The Creutz model~\cite{PhysRevB.99.054302,PhysRevLett.83.2636}, originally proposed by Creutz,  describes the dynamics of spinless fermions on a two-leg ladder system depicted in Fig.~\ref{Figure1}, which is dictated by the Hamiltonian,
\begin{equation}
	\begin{aligned}
		H=-\sum_{n=1}^L & {\left[K\left(e^{i \theta} c_{n+1}^{p \dagger} c_n^p+e^{-i \theta} c_{n+1}^{q \dagger} c_n^q\right)\right.} \\
		& \left.+K\left(c_{n+1}^{p \dagger} c_n^q+c_{n+1}^{q \dagger} c_n^p\right)+M c_n^{q \dagger} c_n^p\right]+ \text { H.c.. }
	\end{aligned}
\end{equation}
Here, the fermionic operators $c_n^p$ and $c_n^q$ are associated with the $n$th site of the lower and upper legs of the ladder, respectively. We assume periodic boundary conditions where $c_{L+1}^{p/q} = c_1^{p/q}$. The hopping amplitudes of the horizontal and diagonal bonds are the same, described by the parameter $K$. The vertical bonds in the system are characterized by the parameter $M$, which is a positive, real value. The ladder is subjected to a magnetic field perpendicular to its surface, which results in the induction of a magnetic flux denoted by the symbol $\theta$.

Let us introduce the spinor $\Gamma^{\dagger}=({c_{k}^{q}}^\dagger {c_{k}^{p}}^\dagger)$, the Hamiltonian after Fourier transformation is written as $H=\sum_{k\ge 0} \Gamma^{\dagger} H(k)\Gamma$, with 
\begin{equation}
	H(k)=-\left(\begin{array}{cc}
		\varepsilon_k^q    & \varepsilon_k^{qp}  \\
		\varepsilon_k^{qp}  &  \varepsilon_k^p
	\end{array}\right),
\end{equation}
where $\varepsilon_k^q =2K\cos(k-\theta)$, $\varepsilon_k^p =2K\cos(k+\theta)$ and $\varepsilon_k^{qp} =2K\cos(k)+M$. The $k$ in momentum space takes discrete values, $k=k_{j}=2 \pi j/L$ with $j=0,1,\dots,L-1$. To obtain a diagonalized Hamiltonian, we use the following transformation~\cite{zhu2016bogoliubov,PhysRevA.80.042315},
\begin{equation}
	\begin{aligned}
		c_{k}^q&=\cos(\gamma_{k}/2)\alpha_{k}+\sin(\gamma_{k}/2)\beta_{k},\\
		c_{k}^p&=-\sin(\gamma_{k}/2)\alpha_{k}+\cos(\gamma_{k}/2)\beta_{k},
	\end{aligned}
\end{equation}

with 
\begin{equation}
	\tan(\gamma_k)=2\varepsilon_k^{qp}/(\varepsilon_k^q-\varepsilon_k^p).\label{EQ5}
\end{equation}
Here $\alpha_{k}$ and $\beta_{k}$ are quasiparticle operators. Then the Hamiltonian can be written as, $H=\sum_k (\varepsilon_k^{\alpha}\alpha_{k}^{\dagger}\alpha_{k}+\varepsilon_k^{\beta}\beta_{k}^{\dagger}\beta_{k})$, where 
\begin{equation}
	\begin{aligned}
		\varepsilon_k^{\alpha}(\theta)&=-2K\cos(k)\cos(\theta)-\sqrt{(\varepsilon_k^{qp})^2+[2K\sin(k)\sin(\theta)]^2},\\
		\varepsilon_k^{\beta}(\theta)&=-2K\cos(k)\cos(\theta)+\sqrt{(\varepsilon_k^{qp})^2+[2K\sin(k)\sin(\theta)]^2},
	\end{aligned}
\end{equation}
with corresponding quasiparticle eigenstates defined by 
\begin{equation}
	\begin{aligned}
		& \alpha_k^{\dagger}|V\rangle=\cos \left(\frac{\gamma_k}{2}\right) {c_k^{q}}^{\dagger}|0\rangle-\sin \left(\frac{\gamma_k}{2}\right) {c_k^{p}}^{\dagger}|0\rangle, \\
		& \beta_k^{\dagger}|V\rangle=\sin \left(\frac{\gamma_k}{2}\right) {c_k^{q}}^{\dagger}|0\rangle+\cos \left(\frac{\gamma_k}{2}\right) {c_k^{p}}^{\dagger}|0\rangle,
	\end{aligned}
\end{equation}
where $|0\rangle$ and $|V\rangle$  are vacuum states of fermion and quasiparticle, respectively. In what follows we will focus on the scenario where the vertical hopping strength $M$ is less than $2K$ and has redefined quasiparticle energies $\widetilde{\varepsilon}_k^{\alpha/\beta}\equiv\varepsilon_{k}^{\alpha/\beta}-M$. In this case it has been observed that the model exhibits second-order quantum phase transitions (QPTs) at $\theta=\theta_{c}=0,\pi$~\cite{PhysRevLett.83.2636,PhysRevB.99.054302,PhysRevLett.102.135702}. The band gap $\Delta\widetilde{\varepsilon}_{k}(\theta_{c})=\widetilde{\varepsilon}_k^\beta (\theta_{c})-\widetilde{\varepsilon}_k^\alpha (\theta_{c})$ at critical point closes at wave numbers $k_c^{\pm}=\pi \pm \arccos(M/2K)$. For a finite system, the wave number $k$ is quantized, and it can only take on discrete values. Suppose we select values of $K$ and $M$ such that $\arccos(M/2K) = (s/r)\pi$ with $s/r \in \mathcal{Q}^+$. In that case, the gap in the system is said to vanish, and this necessitates the number of sites $L$ on each chain to be a multiple of $2r/(r-s)$ and $2r/(r+s)$, i.e.,
\begin{equation}
	L=\frac{2r}{r\pm s}m^{\pm}, m^{\pm} \in \mathbb{N}.\label{EQ8}
\end{equation}
If, when $\arccos(M/2K) = (s/r)\pi$, the system size $L$ does not satisfy these conditions, then the gap will only close asymptotically in the thermodynamic limit. It has been demonstrated that the zero-energy modes that arise when the gap closes hold the key to comprehending the revival structure of the Loschmidt echo after a sudden quench to quantum critical points. However, regardless of whether these conditions governed by the hopping amplitudes are satisfied or not, they have enabled us to uncover the physical mechanism underlying the dynamical behavior of the system. In the following sections, we will demonstrate the pivotal role played by the gap-closing modes in the quench dynamics.
\section{Exact zeros of LE at finite size\label{section3} }	
We will now analyze the quench dynamics of the Creutz model using the LE tool, as described in ~\cite{GORIN200633}. To carry out the quench, we will abruptly change the Peierls phase $\theta_1$ to $\theta_2$ at time $t=0$. We begin by preparing the initial state in the ground state $|\Psi_G (\theta_1)\rangle$, which is given by applying the creation operator $\alpha_{k}^{\dagger}$ to the vacuum state $|V\rangle$ for all negative energy quasiparticle states. We also assume that the Fermi level is selected at zero energy. The LE, after a straightforward calculations yields
\begin{equation}
	\mathcal{L}(t)=\prod_k \mathcal{L}_k (t)=\prod_k \left[1-A_{k}\sin^2 \left(\frac{\Delta\widetilde{\varepsilon}_k t}{2}\right)\right],
\end{equation}
where
\begin{equation}
	\begin{aligned}
		\Delta\widetilde{\varepsilon}_k&=\widetilde{\varepsilon}_{k}^\beta (\theta_2)-\widetilde{\varepsilon}_{k}^\alpha(\theta_2)=2\sqrt{(\varepsilon_{k}^{qp})^2+[2K\sin(k)\sin(\theta_2)]^2},\\
		A_k &=\sin^2 (2\eta_k), 2\eta_k=\gamma_{k}(\theta_1)-\gamma_{k}(\theta_2).\label{EQ10}
	\end{aligned}
\end{equation}
In order to determine the zeros of $\mathcal{L}(t)$, we need to satisfy the condition $A_k=1$. This can be achieved by utilizing equations (\ref{EQ10}) and (\ref{EQ5}), which provide the following explicit constraint condition:
\begin{equation}
	[2K\cos(k)+M]^2=-[2K\sin(k)]^2\sin(\theta_1)\sin(\theta_2).\label{EQ11}
\end{equation}
We should note that the equation is satisfied only when $\sin(\theta_1)\sin(\theta_2)$ is negative semidefinite. This implies that the quench should be performed across or to the critical points, which are located at $\theta_{c}=0$ or $\theta_c=\pi$. Additionally, for a system of finite size, equation (\ref{EQ11}) imposes constraints on the allowed discrete momentum values of $k$. These allowed values of $k$ are therefore determined by the hopping amplitudes and the initial phase $\theta_1$. It is also worth noting that the presence or absence of zero-energy modes will have a significant impact on the dynamical behavior due to the quantization condition related to the system size $L$.

Assuming a pre-quench parameter $\theta_1$ and choosing other parameters $K$ and $M$, we can obtain a series of $\theta_2$ which are determined by Eq.~(\ref{EQ11}) for some discrete value of $k$. Thus, the critical times associated with exact zeros of $\mathcal{L}(t)$ occur when the post-quench parameter $\theta_2$ takes on these discrete values, which are: 
\begin{equation}
	t=t_{n}^{*}=\frac{2\pi}{\Delta\widetilde{\varepsilon}_{k}}(n+\frac{1}{2}),\label{EQ12}
\end{equation}
with 
\begin{equation}
	\Delta\widetilde{\varepsilon}_{k}=2\sqrt{(\varepsilon_{k}^{qp})^2+[2K\sin(k)\sin(\theta_2)]^2}.\label{EQ13}
\end{equation}
We conclude that there exist exact zeros of LE if Eq.~(\ref{EQ11}) is fulfilled. Following the aforementioned analysis of Eq.~(\ref{EQ11}), we shall restrict the initial $\theta_1 \in (0,\pi/2]$, then the semidefinite of $\sin(\theta_1)\sin(\theta_2)$ constrains the post-quench parameter $\theta_2 \in [-\pi/2,0]$, and vice versa. Let us now transform Eq.~(\ref{EQ11}) to
\begin{equation}
	\frac{-[2K\cos(k)+M]^2}{[2K\sin(k)]^2\sin(\theta_1)}=\sin(\theta_2).\label{EQ14}
\end{equation} 
It is crucial to ensure that the absolute value of the left-hand side of the above equation does not exceed one. Given an initial value of $\theta_1$, we can identify a set of wave numbers denoted as $k_s$, which take on discrete values within the discrete Brillouin zone $[0,2(L-1)\pi/L)$. This set of values corresponds to a set of post-quench parameters $\theta_2$, which guarantee that $\mathcal{L}(t)=0$ at the critical times given by Eq.~(\ref{EQ12}). Therefore, it is worthwhile to investigate the system's dynamic behavior within the range of $k_s$, especially near the borders and the zero-energy mode $k_c^{\pm}$. Furthermore, in the thermodynamic limit, the values of $k_s$ are distributed continuously in momentum space.

Let us now consider the case where the allowed values of $k$ are chosen to be the closest to the zero-energy mode. This corresponds to quenching the pre-quench parameter $\theta_1$ to the nearest neighbor of $\theta_{c}=0$. We are interested in studying the shortest distance $\Delta_c(\theta_1)$ between $\theta_{c}=0$ and the solutions of Eq.~(\ref{EQ11}), and how it varies with increasing system size. However, as previously discussed regarding the quantization condition, one must carefully distinguish between the case where the zero-energy modes arise for a finite system size, i.e., $\Delta_c(\theta_1)=0$ for these sizes, and the case where those system sizes do not satisfy the condition.

\begin{figure*}[ht]
	\includegraphics[width=1\linewidth]{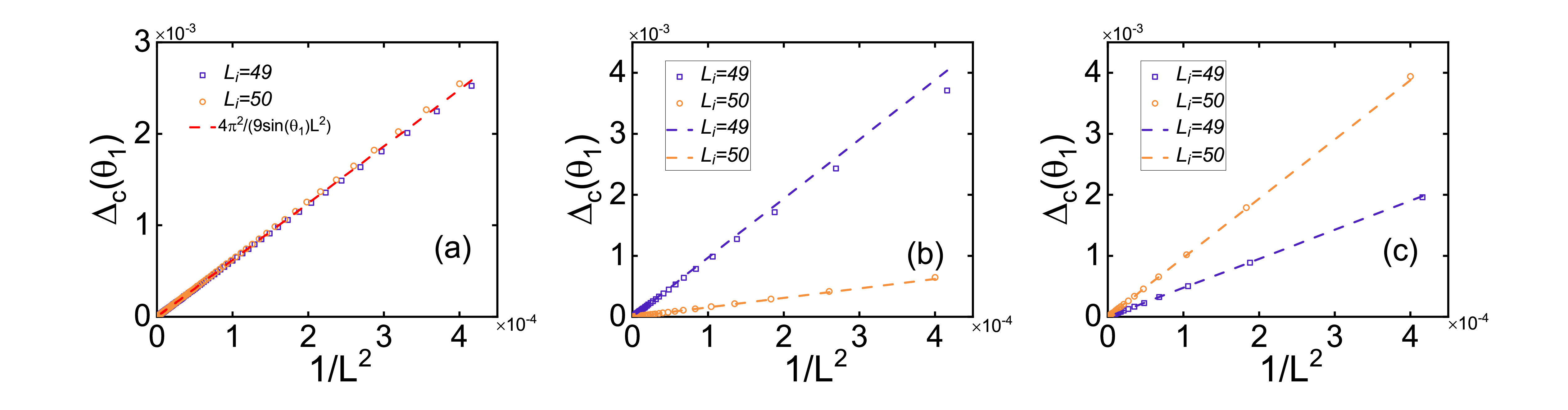}
	
	\caption{The $\Delta_c(\theta_1)$ with respect to $1/L^2$  for the initial hopping parameters chosen as (a) $K=1$, $M=1$. (b) $K=1$, $M=\sqrt{3}$. (c) $K=\sqrt{2}$, $M=-1+\sqrt{3}$. The blue squares and orange circles begin with $L=49$ and $L=50$, respectively. The dashed lines  are corresponding theoretic results. We set the prequench phase as $\theta_1=0.25\pi$.  }
	\label{Figure2}
\end{figure*}
For a finite system with size $L$, the energy gap of the system closes at two specific wave numbers, denoted by $k_c^{\pm}=\pi \pm \arccos(M/2K)$, where $\Delta\widetilde{\varepsilon}_{k_c^{\pm}}=0$. This is only possible if the quantization condition $k_c^{\pm}=2\pi j_c^{\pm}/L$ holds for some integers $j_c^{\pm}$. It is important to note that this is a finite size effect, and for other system sizes which do not satisfy Eq.~(\ref{EQ8}), the gap will not close at these specific wave numbers. In Fig.~\ref{Figure2}, the numerical result of $\Delta_c(\theta_1)$ is shown for $\theta_1=0.25\pi$ with different hopping parameters $K$ and $M$. The choices of system size $L$ depend on the hopping parameters. When we choose $K=M=1$, the gap closes at $k_c^{\pm}=\pi \pm \pi/3$, we thus have $r=3$, $s=1$. If $L$ satisfy the condition Eq.~(\ref{EQ8}), i.e., $L=3m, m\in\mathbb{N}$, in this case $\Delta_c(\theta_1)=0$. The inquiry is how the minimum distance to $\theta_{c}$ behaves as $L$ increases, given that $L \neq 3m$. To address this question, we raise $L$ in steps of three. Two scenarios arise where $L$ fail to satisfy Eq.~(\ref{EQ8}), we choose $L_i=49,50$ as our initial values. As depicted in Fig.~\ref{Figure2}(a), two linear curves with different initial $L$ emerge with distinct slopes which are induced by the nonreciprocal $k$ on both sides of $k_c^{\pm}$. The approximate formula of the $\Delta_c(\theta_1)$ for a large $L$ can be derived from Eq.~(\ref{EQ8}),  and given by 
\begin{equation}
	\Delta_c(\theta_1) \approx \min_{k \in k_s}\frac{(k-k_c^{-})^2}{\sin(\theta_1)},
\end{equation} 
where $k$ is in the region $k_s$ determined by Eq.~(\ref{EQ11}). For the hopping parameters, $K=1$ and $M=1$, we have $k_c^{-}=2\pi/3$. Then $\Delta_c(\theta_1)$ will reduce to ${4\pi^2}/({9\sin(\theta_1)L^2})$ and the fitting line shows good performance when $L$ is large.  

Analogously, we can extend the above-discussed analysis to the scenarios where $K=1$, $M=\sqrt{3}$, and $K=\sqrt{2}$, $M=-1+\sqrt{3}$. In the case of $K=1$, $M=\sqrt{3}$, we expect that $L=12m$ induces the shortest distance $\Delta_c(\theta_1)=0$. Thus, we have $(12-1)$ options for $L$, each corresponding to a different slope of the curves, and we increase $L$ in increments of 12. On the other hand, when $K=\sqrt{2}$ and $M=-1+\sqrt{3}$, $L=24m$ is again expected to give rise to $\Delta_c(\theta_1)=0$. Therefore, we have $(24-1)$ options for $L$, each corresponding to a different slope of the curves, and we increase $L$ in steps of 24. In Fig.~\ref{Figure2}(b) and (c), We consider only the initial values $L_i=49,50$. It can be seen that the fitted curves perform well for large $L$, and curves with different initial $L_i$ have distinct slopes. In the thermodynamic limit, $\Delta_c(\theta_1)$ approaches zero.

\section{Orthogonality catastrophe and quantum speed limit\label{section4}}

We have previously shown that the existence of exact zeros in the LE is a manifestation of OC when the pre- and post-quench parameters, $\theta_1$ and $\theta_2$, meet the condition given in Eq.~(\ref{EQ11}). We have also defined the permitted wave numbers, $k_s$, of the discrete Brillouin zone. However, an investigation into the critical times at which these exact zeros or OC occur can also be intriguing, especially in terms of the intrinsic speed at which the complete system can achieve orthogonality. A relevant measure in this regard is the QSL time~\cite{PhysRevA.67.052109,PhysRevLett.103.160502,PhysRevB.104.094311}, which we will elaborate on in the subsequent text.

\begin{figure}[ht]
	\includegraphics[width=1\linewidth]{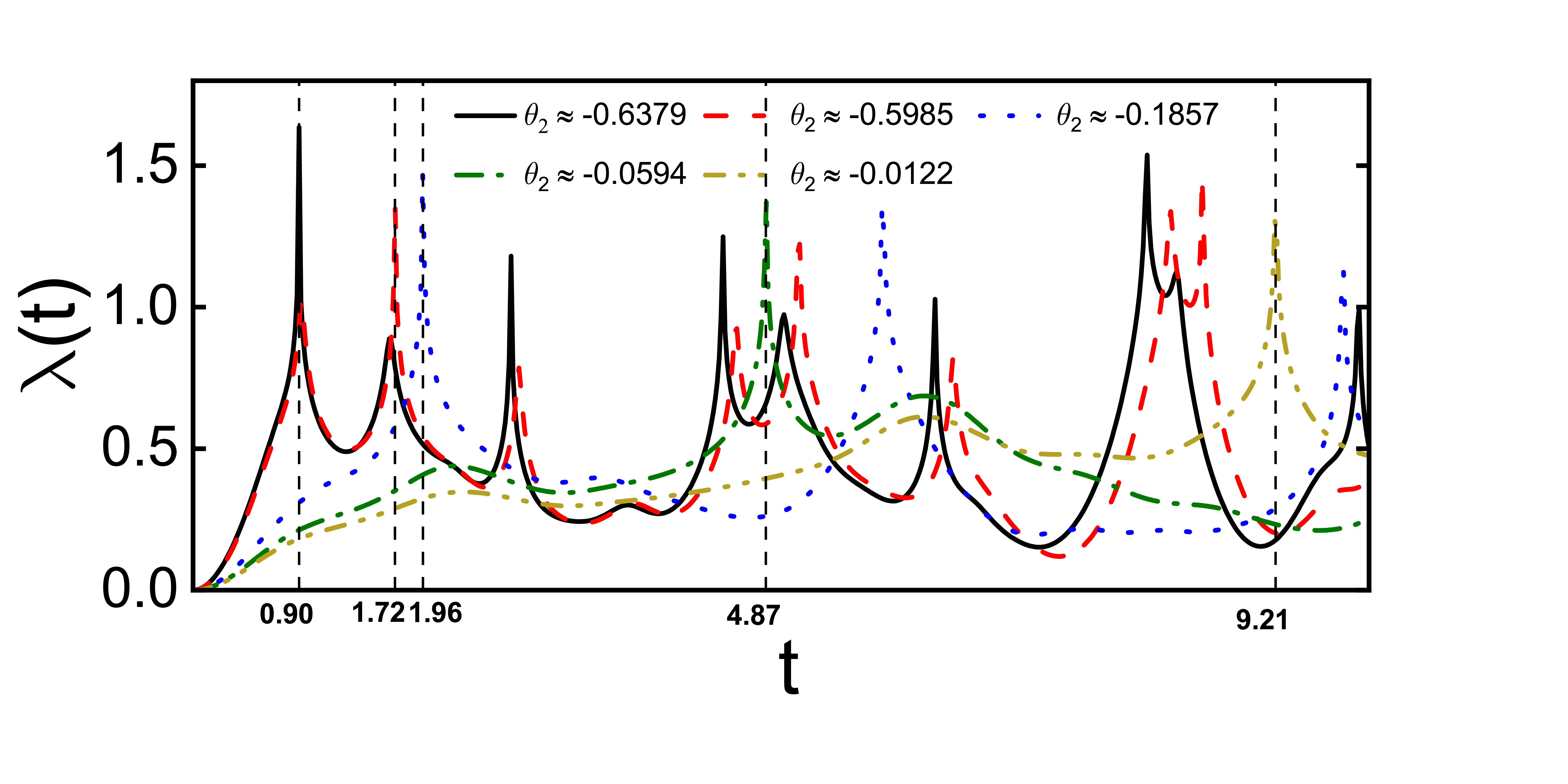}
	\vspace{-0.6cm}
	\caption{Rate function $\lambda(t) $ for $L=22$. Vertical dashed lines represent the first time for the appearance of the exact zero of LE for the allowed $\theta_2$ determined by Eq.~(\ref{EQ11}). The initial parameters are $\theta_1=0.25\pi$, $K=1$, $M=1$.}
	\label{Figure3}
\end{figure}

In Fig.~\ref{Figure3}, we present the rate function $\lambda(t)=-\frac{1}{L}\ln \mathcal{L}(t)$ for a system with $L=22$ and initial phase $\theta_1=0.25\pi$. It can be observed that, for certain values of $\theta_2$, a series of divergence points appear on the real-time axis, which correspond to exact zeros LE and serve as a signature of DQPT. The first divergence time of the rate function, denoted as $\tau_f$ and marked by a black dashed line, represents the minimum critical time required for the system to reach the zeros of LE. $\tau_f$ is calculated as:
\begin{equation}
	\tau_{f}=\frac{\pi}{\Delta\widetilde{\varepsilon}_{k}}.
\end{equation}

\begin{figure}[ht]
	\includegraphics[width=1\linewidth]{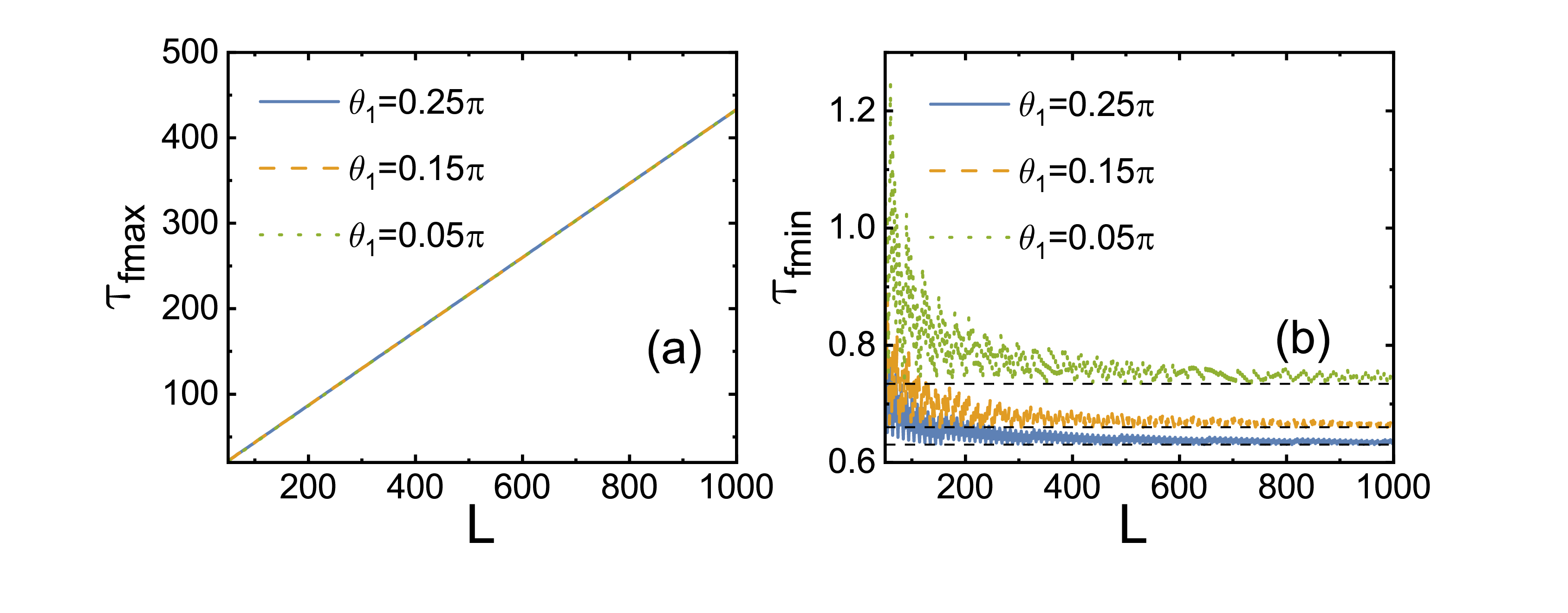}
	
	\caption{(a) $\tau_{fmin}$ and (b)$\tau_{fmax}$ which are determined by Eq.~(\ref{EQ11}) as a function of $L$ for three different initial $\theta_1$. The  hopping paramenters are chosen as $K=1$ and $M=1$. $L$ starts with $L_i$=50. }
	\label{Figure4}
\end{figure}

As depicted in Fig.~\ref{Figure3}, we observe that $\tau_{f}$ increases as $\theta_2$ approaching the critical point $\theta_{c}=0$ and decreases as $\theta_2$ away from the critical point. In particular, we denote the maximal value of $\tau_{f}$ as $\tau_{fmax}=\max [\tau_{f}]$ for various $\theta_2$ determined by Eq.~(\ref{EQ11}). It is found that $\tau_{fmax}$ corresponds to the quench process with $\theta_2$ nearest to $\theta_{c}=0$. As analyzed before and with the form of $\Delta\widetilde{\varepsilon}_{k}$ in Eq.~(\ref{EQ13}), it follows that if the system size $L$ satisfy Eq.~(\ref{EQ8}), $\Delta\widetilde{\varepsilon}_{k}$ equals zero which corresponds to $\theta_2=0$, and an infinite $\tau_{fmax}$. We choose  $L_i=50$, which do no satisfy the condition, as initial values and increase in steps determined by the hopping amplitudes. As displayed in Fig.~\ref{Figure4}(a), the $\tau_{fmax}$ increase linearly with the system size and is independent on the initial phase $\theta_1$. The approximate formula for it can be obtained by simply setting the quenched parameter $\theta_2$ to 0, yielding

\begin{equation}
	\tau_{fmax}=\max_{k \in k_s}\frac{\pi}{2|2K\sin k_c^{-}(k-k_c^{-})|},
\end{equation}
where $k_s$ is determined by Eq.~(\ref{EQ11}) and the wave number $k_c^{-}$ is the gap closing point, i.e., zero energy mode, in the thermodynamic limit. For the case where the hopping parameters are $K=1$ and $M=1$, we have $k_c^{-}={2\pi}/{3}$, then $	\tau_{fmax}=\sqrt{3}L/4$ and it fits well with the numerical results.
We can conclude that as $L$ approaches infinity, $\tau_{fmax}$ also approaches infinity, which corresponds to $\theta_2$ approaching the critical point. Therefore,  we cannot observe the DQPT in a finite time when the system is quenched from the non-critical phase to the critical phase $\theta_2=0$. 
%unless the emerging of zero-energy modes. 

Also, it is interesting to exploit the case when $\theta_2$ determined by Eq.~(\ref{EQ11}) away from the critical $\theta_c=0$. We denote the minimum value of $\tau_{f}$ as $\tau_{fmin}=\min[\tau_{f}]$ for those $\theta_2$. It can be observed that $\tau_{fmin}$ appears around the frontiers of the subregion $k_s$. Alternatively, in the thermodynamic limit, we find the minimum $\tau_{fmin}$ occurs if quenched to $\theta_2=-\pi/2$.  In Fig.~\ref{Figure4}(b), it is demonstrated that $\tau_{fmin}$ initially decrease in an oscillation way with $L$ and eventually converge to certain constant values $\tau_c$. The magnitudes of these constants are dependent on the initial $\theta_1$, hopping parameters $K$, $M$ and have the form,
\begin{equation}
	\tau_c=\frac{\pi}{2|2K\sin k_{c1}|\sqrt{1+{\sin \theta_1}}},
\end{equation} 
where $k_{c1}$ is the solution of Eq.~(\ref{EQ11}) by setting $\theta_2=-\pi/2$, which makes $\tau_c$ smaller. The asymptotic values of the $\tau_c$ for distinct phases $\theta_1$ are depicted as the black dashed lines in Fig.~\ref{Figure4}(b).

\begin{figure}[ht]
	\includegraphics[width=1\linewidth]{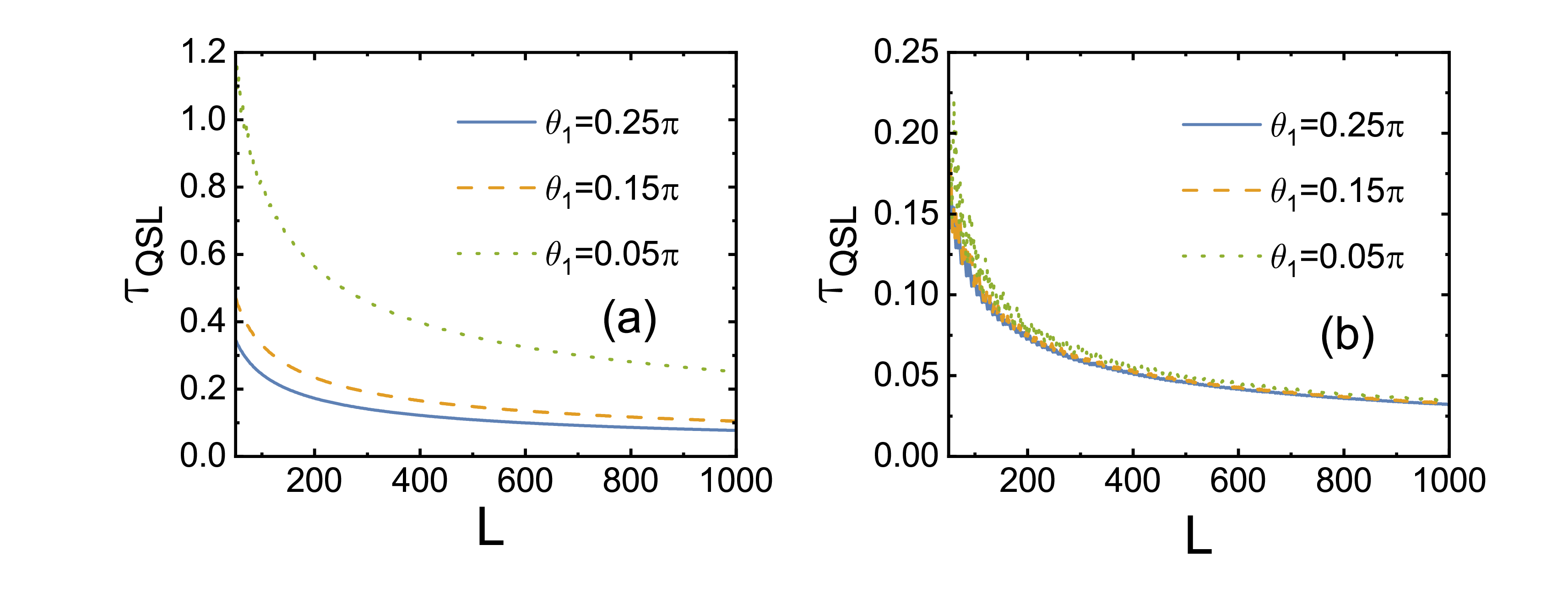}
	
	\caption{$\tau_{QSL} $ vs $L$ for two cases where, (a) $\tau_{f}$ takes its maximum, and (b)$\tau_{f}$ takes its minimum which are determined by Eq.~(\ref{EQ11}).}
	\label{Figure5}
\end{figure}

The QSL sets a bound on the rate of the evolution of system~\cite{Deffner_2017,1992AmJPh..60..182V,PhysRevLett.70.3365,jing2016fundamental} and has been used in exploring various aspects of physics, including quantum state transmission~\cite{PhysRevA.82.022318,PhysRevA.104.052424,PhysRevA.84.022330,doi:10.1080/00107510701342313}, quantum computation~\cite{MARGOLUS1998188}, quantum metrology~\cite{giovannetti2011advances,chen2022enhanced} and quantum control~\cite{PhysRevLett.103.240501,PhysRevA.100.063401}. This bound is given by 
\begin{equation}
	\tau_{QSL}=\frac{\arccos \sqrt{\mathcal{L}(t)}}{\Delta E},
\end{equation}
where $(\Delta E)^2=\langle \Psi_G|H_{\theta_2}^2|\Psi_G\rangle-(\langle \Psi_G|H_{\theta_2}|\Psi_G\rangle)^2$ with $H_{\theta_2}$ represents the post-quench Hamiltonian. When $\mathcal{L}(t)=0$, this bound reduces to the Mandelstam-Tamm bound~\cite{mandelstam1945uncertainty}
\begin{equation}
	\tau_{QSL}=\frac{\pi}{2\Delta E}.
\end{equation} 
Our analysis involves two cases demonstrated before. In the first case, $\tau_{f}$ reaches its maximal value $\tau_{fmax}$, while in the second case, $\tau_{f}$ has its minimum value $\tau_{fmin}$. Let us begin with a general equation for the variance of the Hamiltonian after a sudden quench,  the explicit form of $\Delta E$ which can be deduced from the notation defined for the Creutz model, is given by
\begin{equation}
	(\Delta E)^2=\sum_k \frac{1}{4}\sin^2(2\eta_k)(\widetilde{\varepsilon}_k^{\alpha}(\theta_2)-\widetilde{\varepsilon}_k^{\beta}(\theta_2))^2.
\end{equation} 
In Fig.~\ref{Figure5}, we present the illustration of $\tau_{QSL}$ for both cases, and it is evident that as $L$ increases, $\tau_{QSL}$ diminishes. This phenomenon is due to the fact that the variance of the post-quench Hamiltonian grows as system size $L$ increases. In contrast to the exact results of $\tau_{fmax}$ and $\tau_{fmin}$, we observe significant differences between them, indicating that the bound $\tau_{QSL}$ is not tight. To get a better understanding of this behavior,  we now concentrate on the band gap and its specific form,  $\Delta\widetilde{\varepsilon}_{k_c^-}(\theta_2)$ in Eq.~(\ref{EQ10}), the band gap is finite and has its maximum value for fixed hopping parameters $K$ and $M$, a variation of the post-quench phase $\theta_2$. Then the first divergence time of rate function $\tau_{f}={\pi}/{\Delta\widetilde{\varepsilon}_k(\theta_2)}$ has its minimum value, but would never approach zero. Alternatively, the QSL bound is determined by the energy variance of the post-quench Hamiltonian which takes all the $k$ modes into consideration and is a decreasing function of system size $L$. Therefore, we ascribe the loose bound behavior to the scaling properties of the band gap and the energy variance with respect to the system size. 

\begin{figure}[ht]
	\includegraphics[width=1\linewidth]{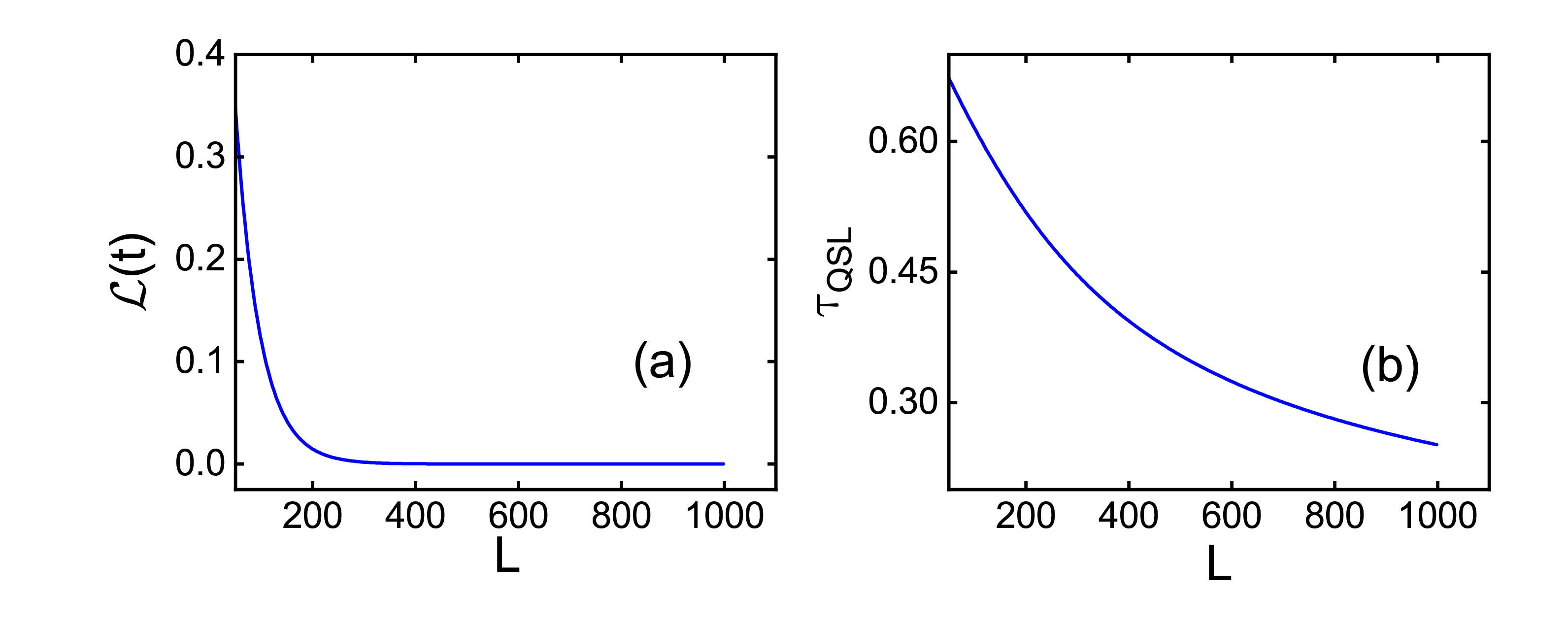}
	
	\caption{(a)$\mathcal{L}(t) $ with driving time $t=1$ and (b)$\tau_{QSL} $ vs $L$ for the case where $L$ satisfy the finite-size quantization condition. We have set initial parameters $\theta_1=0.05\pi$, $K=1$, $M=1$ and post-quench parameter $\theta_2=0$.}
	\label{Figure6}
\end{figure}

However, if we focus on the LE and consider the contributions of all the $k$ modes. It can be found that the decay of LE is sharper with an increase of the system size. As depicted in Fig.~\ref{Figure6}, $\mathcal{L}(t)$ is approaching zero as an increase of $L$ and the corresponding QSL time also decays with $L$. This behavior can be a manifestation of the connection between the QSL and OC~\cite{PhysRevLett.124.110601}.

\begin{figure}[ht]
	
	\includegraphics[width=1\linewidth]{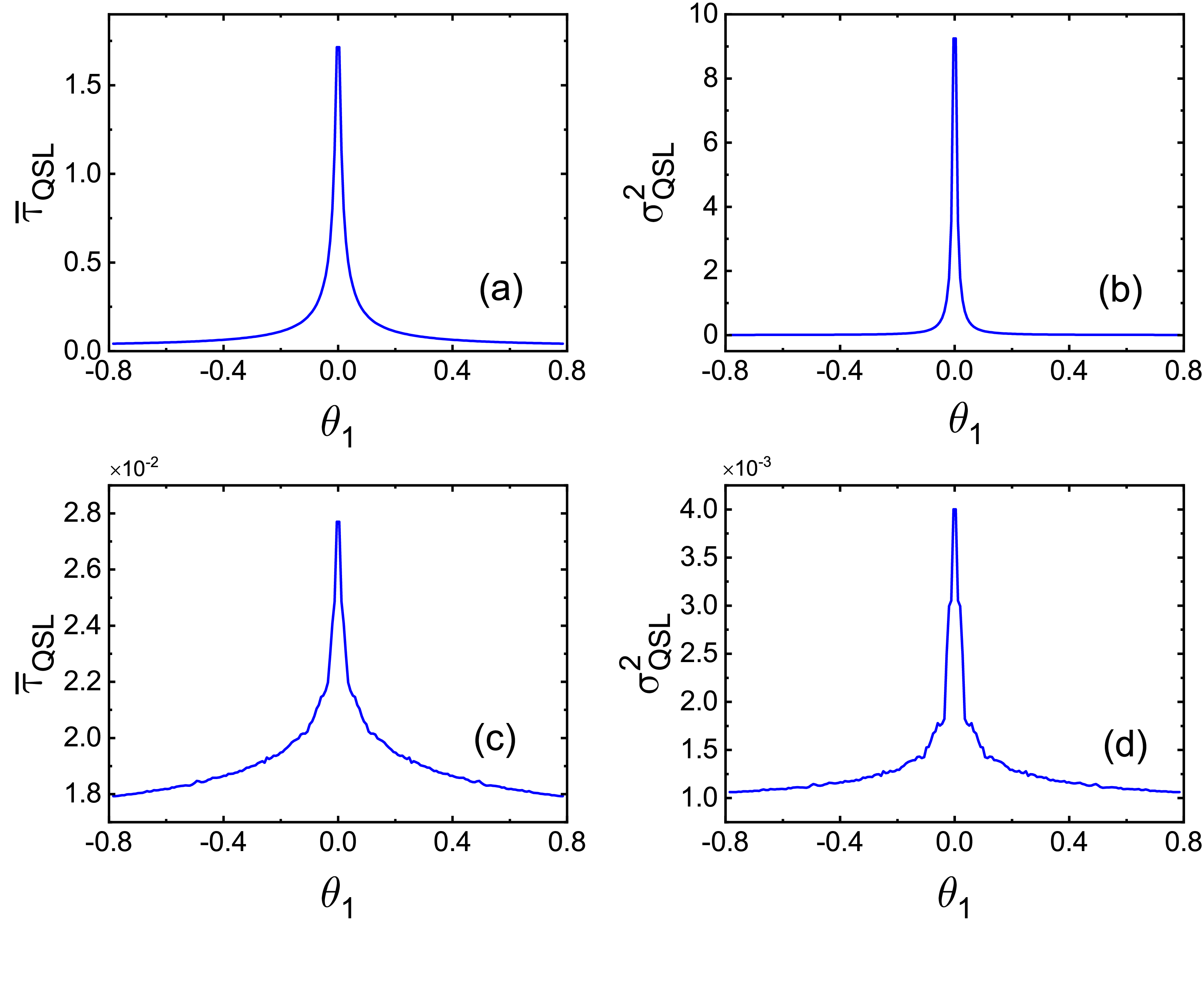}
	
	\caption{$\bar{\tau}_{QSL}$ and $\sigma_{{\tau}_{QSL}}^2$ with  respect to prequench parameter $\theta_1$ for two cases discussed before, where (a) and (b) $\tau_{f} $ takes its maximum, (c) and (d) $\tau_{f} $ takes its minimum.  We have set other initial parameters $K=1$, $M=1$.}
	\label{Figure7}
\end{figure}

Moreover, it has been proposed that the QSL may offer valuable information in identifying the point of static quantum phase transition. We perform numerical calculations of the average value of $\tau_{QSL}(L)$, varying $L$ from $L_{min}$ to $L_{max}$, in order to observe its behavior with respect to $\theta_1$ and illustrate it as: 
\begin{equation}
	\bar{\tau}_{QSL}=\frac{1}{L_{max}-L_{min}}\sum_{L=L_{min}}^{L_{max}}\tau_{QSL}(L).
\end{equation} 
The variance of $\tau_{QSL}(L)$ can be written as:
\begin{equation}
	\sigma_{{\tau}_{QSL}}^2=\frac{1}{L_{max}-L_{min}}\sum_{L=L_{min}}^{L_{max}}[\tau_{QSL}^{2}(L)-\bar{\tau}_{QSL}^2].
\end{equation}
We investigate two scenarios presented in Fig.~\ref{Figure5} and carry out a count from $L_{min}=50$ to $L_{max}=1001$ in increments of three. The numerical results of $\bar{\tau}_{QSL}$ and $\sigma_{{\tau}_{QSL}}^2$ with respect to $\theta_1$ are shown in Fig.~\ref{Figure7}. It is evidenced that both $\bar{\tau}_{QSL}$ and $\sigma_{{\tau}_{QSL}}^2$ exhibit non-analytical behaviors at $\theta_1=0$, which is the static quantum phase transition point in the thermodynamic limit. Hence, we conclude that the QSL holds the potential to function as a useful tool for  indicating the static quantum phase transition.

\section{Noise induced behaviors of quantum speed limit\label{section5}}
We further investigate the impact of noise on the behaviors of QSL.The internal field noise from various sources can be described by random classical noise $\eta$. We assume the noise is applied to the post-quench parameter $\theta_2$, while the initial $\theta_1$ remains constant. The initial state of the system is prepared in the ground state $|\Psi_G(\theta_1)\rangle$. By applying the concept of relative purity\cite{PhysRevLett.110.050403}, the LE for a mixed state can be dictated by 
\begin{equation}
	\mathcal{L}(t)=\frac{\operatorname{tr}[\rho_{0}\rho_{t}]}{\operatorname{tr}(\rho_{0}^2)},
\end{equation} 
where $\rho_{0}=|\Psi_G\rangle \langle \Psi_G|$ and $\rho_{t}$ in the presence of noise is expressed as 
\begin{equation}
	\rho_{t}=\frac{1}{Z}{\sum_{\eta}e^{-iH_{f}^{\eta}t}\rho_{0}e^{iH_{f}^{\eta}t} },
\end{equation}
where $Z=\operatorname{tr}(\sum_{\eta}e^{-iH_{f}^{\eta}t}\rho_{0}e^{iH_{f}^{\eta}t})$ and the sum is taken over all random classical noise. $H_{f}^{\eta}$ represents the post-quench Hamiltonian with noise.

The bound on the rate of the evolution can be derived by using general quantum channels. Note that $\operatorname{tr}(\rho_{0}^2)=1$ for initial pure state and let us parametrize $\mathcal{L}(t)=\cos(\theta)$, we obtain a bound 
\begin{equation}
	\tau_{QSL}=\frac{2\theta^2}{\pi^2}\frac{1}{\overline{\sum_{\eta}||K_{\eta}\rho_{0}\dot{K}_{\eta}^{\dagger}||}},
\end{equation}
where $\overline{X}=\tau_{\theta}^{-1}\int_{0}^{\tau_{\theta}} X dt$,  $||A||=\sqrt{\operatorname{tr}(A^{\dagger}A)}$ and $K_{\eta}=1/\sqrt{Z}\exp[-i(H_{f}^{\eta}-\langle \Psi_G|H_{f}^{\eta}|\Psi_G\rangle )t]$. We hence have
\begin{equation}
	\overline{\sum_{\eta}||K_{\eta}\rho_{0}\dot{K}_{\eta}^{\dagger}||}=\frac{1}{Z}\sum_{\eta}\Delta E^{\eta},
\end{equation}
where $(\Delta E^{\eta})^2=\langle \Psi_G|H_{f}^{\eta 2}|\Psi_G\rangle-(\langle \Psi_G|H_{f}^{\eta}|\Psi_G\rangle)^2$.

\begin{figure}[ht]
	\includegraphics[width=1\linewidth]{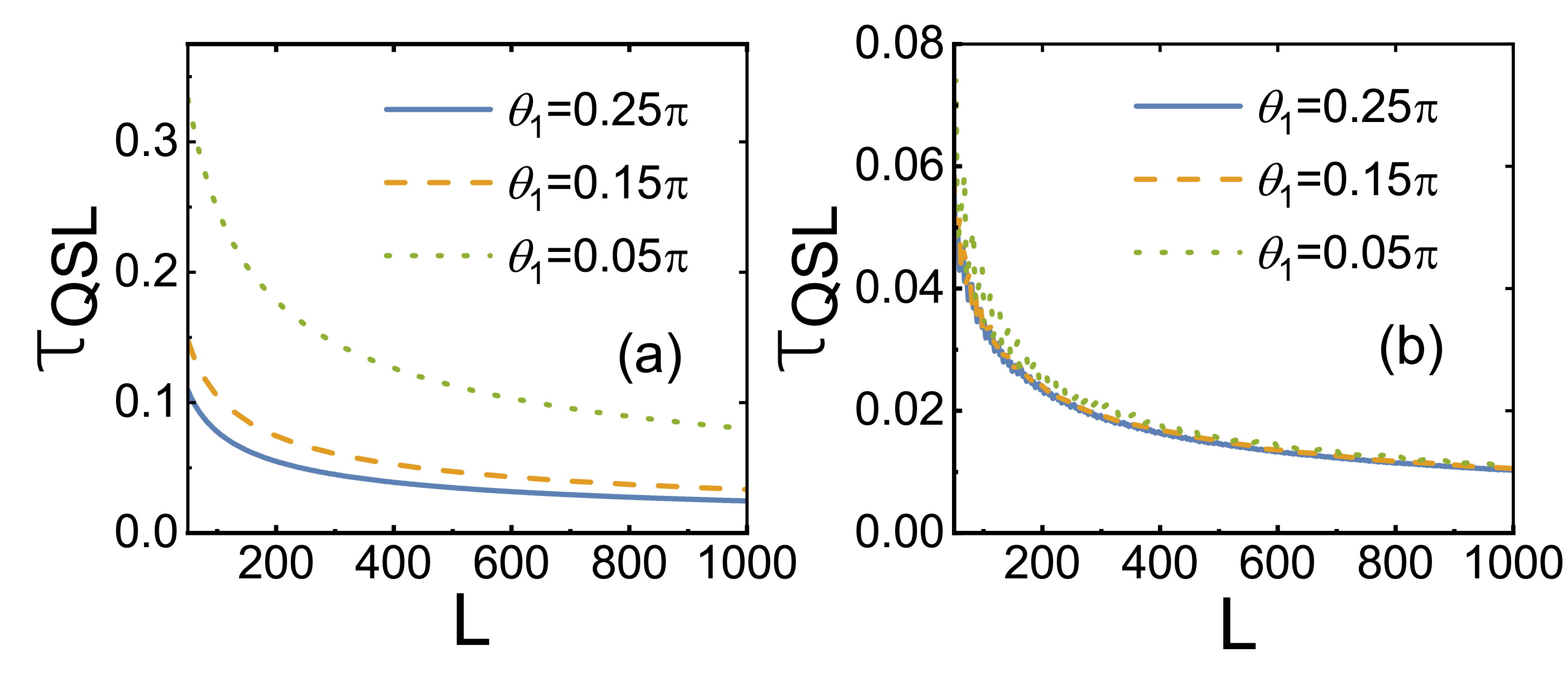}
	
	\caption{Noise induced behaviors of $\tau_{QSL} $ vs $L$ for two cases discussed before, where (a) $\tau_{f}$ takes its maximum, and (b)$\tau_{f}$ takes its minimum which are determined by Eq.~(\ref{EQ11}). We have set the range of the noise $\eta$ in  $\left\{-0.1\theta_2, 0.1\theta_2\right\}$ and the number taken as 1000. }
	\label{Figure8}
\end{figure}

In Figure~\ref{Figure8}, we illustrate the effects of noise on the behaviors of QSL. Two scenarios are shown, and we observe comparable behavior to the noiseless cases, except for a reduced amplitude of the QSL with noise. We maintain that when noise is present, the system's internal minimum time to achieve orthogonality decreases.
\section{Conclusion\label{section6}}
In summary, we investigated the quench dynamics of the Creutz model, which describes the hopping of spinless fermions on a two-leg ladder subjected to a magnetic field. We used the LE to explore the dynamic evolution properties of the system. Our findings show that there exist zeros of LE for finite size $L$ when the post-quench parameter takes discrete values determined by condition Eq.~(\ref{EQ11}). It is worth noting that the zero-energy mode appears at the quantum critical point only when the size $L$ is commensurate with the quantization condition. To highlight the role of the zero-energy mode, we also analyzed the case where $L$ does not satisfy the quantization condition. We found that $\Delta_c(\theta_1)$, defined as the shortest distance between the post-quench parameter $\theta_2$ and the quantum critical point $\theta_{c}=0$, scales linearly with $1/L^2$. Moreover, we observed that different choices of the initial system size result in distinct slopes, which are determined by the zero-energy mode. 

We also investigated the first practical divergence time $\tau_f$ of the rate function for a finite-size system where exact zeros of LE occur. We define the maximum and minimum of $\tau_f$ as $\tau_{fmax}$ and $\tau_{fmin}$, respectively. We found that the maximum of $\tau_f$ grows linearly with $L$ and is associated with the nearest distance to the critical point. However, the minimum of $\tau_f$ converges to a constant value that is not equal to zero as $L$ increases. This suggests that the time when the exact orthogonality of two dynamical evolution states occurs does not vanish in the thermodynamic limit.We calculated the corresponding quantum speed limit time, which is a decreasing function of $L$ for both cases and not tight. This finding seems to contradict the previous knowledge that the time for the two evolving states to reach orthogonality vanishes as $L$ increases and the energy variance scales with $L$. We can explain this discrepancy by noting that the band gap for certain $k$ modes is finite for the system parameters, while the energy variance takes all the $k$ modes into consideration and increases with $L$. We reveal the connection between the quantum speed limit and orthogonality catastrophe by considering the decay of LE for all $k$ modes. Additionally, we have illustrated that the QSL holds the potential to detect the critical point due to the non-analytic behaviors evident in both the mean and variance of $\tau_{QSL}$. Finally, Our demonstration revealed comparable behaviors to the noiseless situation, with the exception of a reduced amplitude of the QSL induced by noise.  
\section*{Acknowledgments}
This work is supported by the National Natural Science Foundation of China under Grants No. 11875086 and No. 11775019. L.-A.W. is supported by the Basque Country Government (Grant No. IT1470-22) and Grant No. PGC2018-101355-B-I00 funded by MCIN/AEI/10.13039/501100011033.
\bibliographystyle{quantum}
\bibliography{Creutzreference.bib}

\nocite{*}

\end{document}